\documentclass[paper]{mystyle}

\usepackage{cite}
\usepackage{epsfig}

\newcommand{\beq}{\begin{equation}}
\newcommand{\eeq}{\end{equation}}
\newcommand{\bqa}{\begin{eqnarray}}
\newcommand{\eqa}{\end{eqnarray}}
\newcommand{\nl}{\nonumber \\}
\def\db#1{\bar D_{#1}}
\def\d#1{D_{#1}}
\def\tld#1{\tilde {#1}}
\def\slh#1{\rlap / {#1}}
\def\eqn#1{Eq.~(\ref{#1})}
\def\eqns#1#2{Eqs.~(\ref{#1}) and~(\ref{#2})}
\def\eqnss#1#2{Eqs.~(\ref{#1})-(\ref{#2})}

\def\app#1{Appendix~\ref{#1}}

\newcommand\fverb{\setbox\pippobox=\hbox\bgroup\verb}
\newcommand\fverbdo{\egroup\medskip\noindent%
                        \fbox{\unhbox\pippobox}\ }
\newcommand\fverbit{\egroup\item[\fbox{\unhbox\pippobox}]}
\newbox\pippobox

\def\spa#1.#2{\left\langle#1\,#2\right\rangle}
\def\spb#1.#2{\left[#1\,#2\right]}

\def\feynsl#1{
  \setbox0=\hbox{/} \setbox1=\hbox{$#1$}
  \dimen0=\wd0 \advance\dimen0 by -\wd1 \divide\dimen0 by 2
  \ifdim\wd0>\wd1 \raise.15ex\copy0\kern-\wd0\kern\dimen0\copy1\kern\dimen0
  \else \kern-\dimen0\raise.15ex\copy0\kern-\dimen0\kern-\wd1\copy1\fi}

\newskip\humongous \humongous=0pt plus 1000pt minus 100pt

\newif\ifdtup

\def    \br(#1,#2)          {\mbox{$\langle #1 \, #2 \rangle$}}
\def    \sq(#1,#2)          {\mbox{$\left[  #1 \, #2 \right]$}}

\title{Reducing full one-loop amplitudes to scalar integrals
       at the integrand level}

\author{Giovanni Ossola\footnote{
e-mail: ossola@inp.demokritos.gr} , Costas G.
Papadopoulos\footnote{e-mail: costas.papadopoulos@cern.ch}\, and
 Roberto Pittau\footnote{Permanent address: Dipartimento di Fisica
Teorica, Univ. di Torino and INFN, sez. di Torino,
Italy.}~\footnote{ e-mail: roberto.pittau@to.infn.it} \\ Institute
of Nuclear Physics, NCSR "DEMOKRITOS", 15310 Athens, Greece.}

\abstract{We show how to extract the coefficients of the 4-, 3-, 2-
and 1-point one-loop scalar integrals from the full one-loop
amplitude of arbitrary scattering processes. In a similar fashion,
also the rational terms can be derived. Basically no information on
the analytical structure of the amplitude is required, making our
method appealing for an efficient numerical implementation.}

\preprint{}
\begin{document}

\section{Introduction}
With the ongoing evolution of the experimental programs of the LHC
and the International Linear Collider, high precision predictions
for multi-particle processes are urgently needed. In the last years
we have seen a remarkable progress in the theoretical description of
multi-particle processes at tree-order, thanks to very efficient
recursive algorithms~\cite{LO}. Nevertheless the current need of
precision goes beyond tree order and therefore a similar description
at the one loop level is more than desirable.

The computation of the one-loop matrix elements seems to be
notoriously difficult. The development of the main ingredients to
accomplish this task, started almost 30 years ago with the
pioneering works of 't Hooft and Veltman~\cite{'tHooft:1978xw} and
Passarino and Veltman~\cite{Passarino:1978jh}. Still nowadays very
few {\it complete} calculations with more than 3
particles\cite{Denner:2005es} in the final state exist \footnote{For
a review of reduction methods, see~\cite{Denner:2005nn}.}.

The problem arises because of two reasons: the complexity of Feynman
graph representation at the one-loop level and the way the reduction
of $n$-point one-loop integrals in terms of scalar 4-, 3-, 2- and
1-point functions is performed. For the former, recursive equations
seem to be a very promising tool. For the latter, it would be
desirable to have a reduction of the full (sub-)amplitude instead of
the individual tensor one-loop integrals. In fact, a method that
will be based on the minimum possible analytical information about
the one-loop amplitude will be more adequate, {\it in principle},
for an efficient numerical implementation\footnote{For other
attempts towards a direct numerical implementation see also
\cite{Ferroglia:2002mz}.}.

During the last years we have seen some very interesting
developments. In the front of the reduction of tensor integrals a
new method at the integrand level has been worked out by Pittau, and
del Aguila~\cite{delAguila:2004nf}. This method will be the starting
point of our work. On the other hand the idea of cut
constructibility~\cite{Bern:1994cg} in computing one-loop amplitudes
(and not just integrals), has been proven very efficient in getting
many very well known results. In the last two years a new
development in computing one-loop amplitudes has been initiated by
the work of Britto, Cachazo and Feng,~\cite{Britto:2004nc} and
combined with the unitarity method, resulted to a series of
remarkable results concerning QCD one-loop multi-particle
amplitudes~\cite{Berger:2006vq}. Moreover the introduction of
quadruple cuts allowed a simplified algebraic approach, at least for
the coefficient of the box function.

In this paper we propose a reduction of arbitrary one-loop
(sub-)amplitude at {\it the integrand level} by exploiting the set
of kinematical equations for the integration
momentum~\cite{kallen:1965}, that extend the already used quadruple,
triple and double cuts. In contrast to the usual method of
cut-contractibility, it is also possible to reconstruct the full
rational terms of the amplitude. The method requires a minimal
information about the form of the one-loop (sub-)amplitude and
therefore it is well suited for a numerical implementation. It also
gives rise to very interesting simplifications of well known
results.

In \cite{delAguila:2004nf} it has been shown, by explicitly reconstructing
denominators, that the {\em integrand} of any $m$-point one-loop amplitude
can be rewritten as
\footnote{We assume $p_0 \ne 0$, for that choice
allows a completely symmetric treatment of all denominators $\db{i}$.}
\bqa
\label{eq:1}
A(\bar q)= \frac{N(q)}{\db{0}\db{1}\cdots \db{m-1}}\,,~~~
\db{i} = ({\bar q} + p_i)^2-m_i^2\,,~~~ p_0 \ne 0\,,
\eqa
where we use a bar to denote objects living
in $n=~4+\epsilon$  dimensions and
where the numerator $N(q)$ can be cast in the form
\bqa
\label{eq:2}
N(q) &=&
\sum_{i_0 < i_1 < i_2 < i_3}^{m-1}
\left[
          d( i_0 i_1 i_2 i_3 ) +
     \tld{d}(q;i_0 i_1 i_2 i_3)
\right]
\prod_{i \ne i_0, i_1, i_2, i_3}^{m-1} \db{i} \nl
     &+&
\sum_{i_0 < i_1 < i_2 }^{m-1}
\left[
          c( i_0 i_1 i_2) +
     \tld{c}(q;i_0 i_1 i_2)
\right]
\prod_{i \ne i_0, i_1, i_2}^{m-1} \db{i} \nl
     &+&
\sum_{i_0 < i_1 }^{m-1}
\left[
          b(i_0 i_1) +
     \tld{b}(q;i_0 i_1)
\right]
\prod_{i \ne i_0, i_1}^{m-1} \db{i} \nl
     &+&
\sum_{i_0}^{m-1}
\left[
          a(i_0) +
     \tld{a}(q;i_0)
\right]
\prod_{i \ne i_0}^{m-1} \db{i} \nl
     &+& \tld{P}(q)
\prod_{i}^{m-1} \db{i}\,. \eqa The quantities $d(i_0 i_1 i_2 i_3)$
are the coefficients of the 4-point loop functions with the four
denominators $\db{i_0} \db{i_1} \db{i_2} \db{i_3}$. Analogously, the
$c(i_0 i_1 i_2)$, $b(i_0 i_1)$ and $a(i_0)$ are the coefficients of
all possible 3-point, 2-point and 1-point loop functions,
respectively.

The ``spurious'' terms $\tld{d}$, $\tld{c}$, $\tld{b}$, $\tld{a}$
and $\tld{P}$ still depend on $q$. They are defined by
the requirement that they should vanish upon integration over $d^n \bar q$,
as we shall see later.

Notice that no coefficient of scalar functions with more that four denominators
appear. This is due to the fact that scalar functions with
$m > 4$ can always be expressed as a linear
combination of 4-point functions and, possibly,
extra $\tld{d}$ terms \cite{vanNeerven:1983vr}.

All $q$'s in the numerator $N(q)$ of \eqn{eq:2} are
4-dimensional. If $n$-dimensional $q$'s
are needed, they can be split into 4-dimensional
and $\epsilon$-dimensional parts
as explained in \cite{delAguila:2004nf}.
\eqn{eq:2} is then applicable to the purely 4-dimensional terms,
that are usually the most difficult to compute.

Since the scalar 1-, 2-, 3-, 4-point functions are known, the only
knowledge of the existence of the decomposition of \eqn{eq:2} allows
one to reduce the problem of calculating $A(\bar q)$ to the
algebraical problem of extracting all possible coefficients in
\eqn{eq:2} by computing $N(q)$ a sufficient number of times, at
different values of $q$, and then inverting the system.

 In carrying out this program two problems arise.
First the explicit knowledge of the spurious terms is needed,
secondly the size of the system should be kept manageable.
 To illustrate this second point, for $m=6$ and without counting any
spurious term, there are 56 independent scalar loop functions
and it is clearly advisable to avoid the inversion of a $56 \times 56$ matrix.
Our solution to this is basically singling out
particular choices of $q$ such that, systematically,
4, 3, 2 or 1 among all possible denominators $\db{i}$ vanishes.
 Then, as we shall see in Section \ref{sec:4}, the system
of equations becomes ``triangular'': first one solves for all
possible 4-point functions, then for the 3-point functions and so
on.

 Notice that the described procedure can be performed
{\em at the amplitude level}. One does not need to
repeat the work for all Feynman diagrams, provided their sum is known.
This circumstance is particularly appealing when
our method is used together with some recursion relation to build up $N(q)$.
 We postpone this problem to a future publication and, in this paper, we
suppose to know $N(q)$.

 A last comment is in order.
 In reconstructing the denominators, there is a mismatch between
the 4-dimensional $q$ in $N(q)$ and the $n$-dimensional denominators
$\db{i}$. To compensate for this it suffices to replace $m_i^2 \to
m_i^2 - \tld{q}^2$ in all the coefficient of \eqn{eq:2}, where
$\tld{q}^2$ is the ($n-4$)-dimensional part of $\bar{q}^2$
\cite{delAguila:2004nf}. The coefficients of the various powers of
$\tld{q}^2$, obtained through this mass shift, are the coefficients
of the extra integrals introduced in \cite{delAguila:2004nf}, which
give rise to nothing but the rational part of the amplitude
\cite{Xiao:2006vr}.

  In this paper we show how to extract,
with the help of \eqn{eq:2}, the coefficients of the loop functions,
including the rational terms, from any amplitude. In Section
\ref{sec:3} we list and compute the needed spurious terms, in
Section \ref{sec:4} we show how to get, in a systematic fashion, the
coefficients of all 4-, 3-, 2- and 1-point scalar integrals, while
Section \ref{sec:5} deals with the problem of adding the missing
rational terms. Finally, in Section \ref{appli}, we present some
practical applications and tests we made on our method.

\section{\label{sec:3} The spurious terms}
Before any attempt of extracting $d(i_0 i_1 i_2 i_3)$,
$c(i_0 i_1 i_2)$, $b(i_0 i_1)$ and $a(i_0)$,
one should explicitly know the $q$ dependence
of the spurious terms $\tld{d}(q;i_0 i_1 i_2 i_3)$,
 $\tld{c}(q;i_0 i_1 i_2)$, $\tld{b}(q;i_0 i_1)$,
$\tld{a}(q;i_0)$ and $\tld{P}(q)$. This can be achieved by
decomposing any 4-dimensional $q$ appearing in the numerator of
\eqn{eq:1} in terms of a convenient basis of massless 4-momenta
\cite{delAguila:2004nf}, the coefficients of which either
reconstruct denominators or vanish upon integration over $d^n \bar
q$. The first terms gives rise to the $d$, $c$, $b$ and $a$
coefficients in \eqn{eq:2}. The second ones to all the additional
spurious terms. The latter category is further classified, in
\eqn{eq:2}, according to the number of the remaining denominators.
In the following, we shall call the $\tld{d}$, $\tld{c}$, $\tld{b}$,
$\tld{a}$, $\tld{P}$ terms  4,3,2,1,0-point like spurious terms,
respectively. As we shall see, the actual form of them generally
depends on the maximum possible rank of the loop tensors in the
amplitude.

 Since the decomposition makes use of
the momenta appearing in the denominators we set, for simplicity,
$i_0= 0$, $i_1= 1$, $i_2= 2$ and $i_3= 3$ and derive explicit
formulae for $\tld{d}(q;0123)$, $\tld{c}(q;012)$, $\tld{b}(q;01)$,
$\tld{a}(q;0)$ and $\tld{P}(q)$. By relabeling the indices, all the
other spurious terms are easily derived. Before carrying out this
program, we recall some basic results of Ref.
\cite{delAguila:2004nf}, also with the aim to set up our notation.

\noindent The explicit decomposition reads
\bqa
\label{eq:4a}
q^\mu &=& -p_0^\mu + \frac{\beta}{\gamma} D^\mu -\frac{1}{2\gamma} Q^\mu\,, \nl
D^\mu &=& \frac{1}{\beta}[
 2 [(q + p_0)\cdot \ell_1]\ell_{2}^\mu
+2 [(q + p_0)\cdot \ell_2]\ell_{1}^\mu ]\,, \nl Q^\mu &=&  [(q+
p_0)\cdot \ell_3] \ell_{4}^\mu
       +   [(q+ p_0)\cdot \ell_4] \ell_{3}^\mu\,,
\eqa where $\ell_1$ and $\ell_2$  are massless 4-vector satisfying
the relations
\bqa \label{eq:4b} k_1   = \ell_1 + \alpha_1
\ell_2\,,~~~k_2   = \ell_2 + \alpha_2 \ell_1 \,, \eqa with \bqa
\label{eq:4bb}
 k_i = p_i-p_0\,.
\eqa
Furthermore, in spinorial notation,
\bqa
\label{eq:4c}
\ell_3^\mu = <\ell_1| \gamma^\mu | \ell_2]\,,~~
\ell_4^\mu = <\ell_2| \gamma^\mu | \ell_1]\,
~~{\rm with}~~ (\ell_3 \cdot \ell_4) = -4(\ell_1 \cdot \ell_2)\,.
\eqa
The solution to \eqn{eq:4b} reads
\bqa
\label{eq:4d}
\ell_1 &=& \beta(k_1- \alpha_1 k_2)\,,~~~ \ell_2
~=~ \beta(k_2- \alpha_2 k_1)\,,\nl
\beta  &=& 1/(1- \alpha_1 \alpha_2)\,,~~~ \alpha_i~=~\frac{k_i^2}{\gamma}\,,\nl
\gamma &\equiv& 2 (\ell_1 \cdot \ell_2) = (k_1 \cdot k_2) \pm \sqrt{\Delta}
\,,~~~\Delta ~=~  (k_1 \cdot k_2)^2-k_1^2 k_2^2\,.
\eqa

A last comment is in order. We make the assumption,
always realized in practical calculations,
to compute the amplitude $A(\bar q)$ in a gauge where the maximum
rank of the appearing loop tensors is never higher
than the number of denominators, as it happens, for example, in the
renormalizable gauge.
This choice limits the number of the spurious terms.
For instance, there is no $\tld{P}$ in such a case.
However, the fact that all gauges are equivalent, leads us to the following
conjecture

\vspace{0.5cm}

{\em One can always limit her/himself to the spurious terms
appearing in the renoramalizable gauge, because, in physical
amplitudes, the contributions coming  from the tensors of higher
rank should add up to zero}.

\vspace{0.5cm}

\noindent This conjecture, being rather strong, has to be checked in
practical calculations. We are now ready to derive the $q$
dependence of the spurious terms.

\subsection{The 4-point like spurious term}
By iteratively using \eqn{eq:4a}, only one possible integrand that vanishes
upon integration is left, namely
\bqa
\label{eq:3a}
\tld{d}(q;0123) = \tld{d}(0123)\, T(q)\,,
\eqa
where $\tld{d}(0123)$ is a constant (namely
does not depend on $q$) and
\bqa
\label{eq:4}
 T(q) \equiv Tr[(\slh{q}+\slh{p_0}) \slh{\ell_1} \slh{\ell_2} \slh{k_3}
\gamma_5]\,.
\eqa
To prove this statement, let us call $N^{(3)}(q)$ the numerator of a term
containing the four denominators $\db{0}\db{1}\db{2}\db{3}$.
$N^{(3)}(q)$ is necessarily a polynomial in $q$, whose
maximum degree we will denote by $j_{max}$. Notice that $j_{max}$ is also
the maximum rank of the 4-point tensors that appear when performing
a standard reduction procedure.
Being interested in terms in which the four original denominators
are not canceled out by the numerator function,
we can systematically neglect all denominators
that are reconstructed from $N^{(3)}(q)$
\footnote{We will also ignore terms proportional to powers of $\tld{q}^2$,
for they give rise to rational parts in the amplitude that can be
treated separately (see Section \ref{sec:5}).}.
In particular, by expressing back $\ell_1$ and $\ell_2$ in terms of
$k_1$ and $k_2$, as shown in \eqn{eq:4d}, one obtains \cite{delAguila:2004nf}
\bqa
\label{eq:4s}
D^\mu &=&
 \frac{2}{\beta} (p_0\cdot \ell_1) \ell_2^\mu
+\frac{2}{\beta} (p_0\cdot \ell_2) \ell_1^\mu
+(\db{1}-\db{0} + h_1)\,r_2^\mu
+(\db{2}-\db{0} + h_2)\,r_1^\mu\,, \nl
 h_i &=& (m_i^2-p_i^2)-(m_0^2-p_0^2)\,,~~~
 r_1^\mu ~=~ \ell_1^\mu -\alpha_1 \ell_2^\mu\,,~~~
 r_2^\mu ~=~ \ell_2^\mu -\alpha_2 \ell_1^\mu\,.
\eqa
Therefore
\bqa
\label{eq:4aa}
D^\mu &=& F^{\mu} + \sum_{i=0}^2{\cal O}(\db{i})\,,
\eqa
where
\bqa \label{eq:4ff}
F^\mu &\equiv&
 \frac{2}{\beta} (p_0\cdot \ell_1) \ell_2^\mu
+\frac{2}{\beta} (p_0\cdot \ell_2) \ell_1^\mu
+h_1\,r_2^\mu + h_2\,r_1^\mu\,,
\eqa
so that
\bqa
\label{eq:4aaa}
q^\mu &=& -p_0^\mu + \frac{\beta}{\gamma} F^\mu
-\frac{1}{2\gamma} Q^\mu
 + \sum_{i=0}^2{\cal O}(\db{i})\,.
\eqa
We used the notation ${\cal O}(\db{i})$ to indicate terms in which
one of the denominators $\db{i}$ has been explicitly reconstructed,
and can therefore be neglected, as far as the construction of spurious
terms is concerned.
By replacing each $q$ appearing in $N^{(3)}(q)$ by the r.h.s. of \eqn{eq:4aaa},
the only possible numerators of degree $j_{max}$, which preserve all
four denominators, turn out to be
\bqa
\label{eq:3terms}
[(q+p_0) \cdot \ell_3]^{j_{max}}\,,~~~
[(q+p_0) \cdot \ell_4]^{j_{max}}\,,~~~
[(q+p_0) \cdot \ell_3]^{j_3} [(q+p_0) \cdot \ell_4]^{j_4}\,,
\eqa
with $j_3+ j_4= j_{max}$.
The rank of such terms
can be reduced with the help of the two following identities
\footnote{A demonstration can be found in
\cite{delAguila:2004nf}. Notice that the factor 2 in front of the last term
of the second equation is missing in that paper.}:
\bqa
\label{eq:4aaaa}
[(q+p_0) \cdot \ell_3] [(q+p_0) \cdot \ell_4] &=&
 \beta (q+p_0)^\alpha D_\alpha -\gamma (q+p_0)^2\,,\nl
{[{(q+p_0)} \cdot {\ell_{3(4)}}]}  [(q+p_0) \cdot \ell_{3(4)}]    &=&
\frac{1}{(k_3 \cdot \ell_{4(3)})}
\left\{ [\gamma (q+p_0)^2 - \beta (q+p_0)^\alpha D_\alpha]
(k_3 \cdot \ell_{3(4)}) \right. \nl
&-& \left.  2 \left[\gamma [(q+p_0)\cdot k_3] - \beta k_3^\alpha D_\alpha
\right]
[(q+p_0)\cdot \ell_{3(4)}]
\right\}\,.
\eqa
In fact, the insertion of \eqns{eq:4aa}{eq:4aaa}
in \eqn{eq:4aaaa},
together with the identities
\bqa
\label{eq:recden}
(q+p_0)^2 ~=~ \db{0}+m_0^2-\tld{q}^2\,,~~~
2\,(q \cdot k_3) ~=~ \db{3}-\db{0} + h_3\,,
\eqa
with $h_3$ given in \eqn{eq:4s}, gives
\bqa
\label{eq:4sa}
[(q+p_0) \cdot \ell_3] [(q+p_0) \cdot \ell_4] &=&
\frac{2 \beta^2 F^2- \beta F \cdot Q}{2 \gamma}
-\gamma\,m_0^2 + \sum_{i=0}^2{\cal O}(\db{i})
+ {\cal O}(\tld{q}^2)\nl
{[{(q+p_0)} \cdot {\ell_{3(4)}}]}  [(q+p_0) \cdot \ell_{3(4)}]    &=&
\frac{1}{(k_3 \cdot \ell_{4(3)})}
\left\{ \left[\gamma\, m_0^2 -
\frac{2 \beta^2 F^2- \beta F \cdot Q}{2 \gamma}\right]
(k_3 \cdot \ell_{3(4)}) \right. \nl
&-& \left.  2 \left[\gamma \left(p_0\cdot k_3 + \frac{h_3}{2}\right)
- \beta \,k_3\cdot F
\right] [(q+p_0)\cdot \ell_{3(4)}]
\right\} \nl
&+& \sum_{i=0}^3{\cal O}(\db{i})
+ {\cal O}(\tld{q}^2)\,.
\eqa
The two equations (\ref{eq:4sa}), when introduced in the numerators
of \eqn{eq:3terms},
reduce their rank
from $j_{max}$ to $j_{max}-1$,
up to contributions containing less denominators
or proportional to $\tld{q}^2$.
By applying the described procedure $j_{max}-1$ times
$N^{(3)}(q)$ takes the form
\bqa
N^{(3)}(q)= \eta_0
+ \eta_3\,[(q+p_0) \cdot \ell_3]
+ \eta_4\,[(q+p_0) \cdot \ell_4]
+ \sum_{i=0}^3{\cal O}(\db{i})
+ {\cal O}(\tld{q}^2)\,,
\eqa
where the coefficients $\eta_0$, $\eta_3$ and $\eta_4$ do not depend on $q$.
The denominators still hidden in $[(q+p_0) \cdot \ell_{3,(4)}]$
can be further extracted by using the identity \cite{delAguila:2004nf}
\bqa
[(q+p_0) \cdot \ell_{3,4}] &=&
\frac{1}{(k_3 \cdot \ell_{4,3})}
\left\{
\beta\, k_3^\alpha D_\alpha - \gamma [(q+p_0)\cdot k_3]
\pm \frac{T(q)}{2}
\right\}\,\nl
&=&
\frac{1}{(k_3 \cdot \ell_{4,3})}
\left\{
\beta\, k_3\cdot F - \gamma \left[p_0\cdot k_3 + \frac{h_3}{2}\right]
\pm \frac{T(q)}{2}
\right\} \nl
&+& \sum_{i=0}^3{\cal O}(\db{i})
+ {\cal O}(\tld{q}^2)\,,
\eqa
with $T(q)$ given in \eqn{eq:4}. Therefore the final
expression for $N^{(3)}(q)$
reads
\bqa
N^{(3)}(q)= d(0123) + \tld{d}(0123)\, T(q)
+ \sum_{i=0}^3{\cal O}(\db{i})
+ {\cal O}(\tld{q}^2)\,.
\eqa
The statement that $\tld{d}(q;0123)$ must have
the form given in \eqn{eq:3a}
is then equivalent to the

\noindent \underline{Theorem}:
\bqa
 \int d^n \bar q \frac{T(q)}{\db{0}\db{1}\db{2}\db{3}} = 0\,.
\eqa
The proof trivially follows by making
the shift $q \to q-p_0$ in the integration
variable and by noticing that $T(q) \propto \epsilon(q,\ell_1,\ell_2,k_3)$.
In fact, the resulting rank one 4-point function can only be proportional
to $k_1$, $k_2$ and $k_3$ and each term vanishes when contracted with the
$\epsilon$ tensor.

\subsection{The 3-point like spurious terms}
To derive $\tld{c}(q;012)$ we make use of the

\noindent \underline{Theorems}:
\bqa
\label{eq:5}
 \int d^n \bar q \frac{[(q+p_0)\cdot \ell_3]^j}{\db{0}\db{1}\db{2}}
  &=& 0\,,\nl
 \int d^n \bar q \frac{[(q+p_0)\cdot \ell_4]^j}{\db{0}\db{1}\db{2}} &=& 0\,,
~~~~\forall j= 1,2,3,\cdots
\eqa Once again, they are proven by shifting $q$ and performing a tensor
decomposition.

  Such integrals come from
the only possible terms in the decomposition of 3-point tensor {\em
integrands} that do not reconstruct denominators
\cite{delAguila:2004nf}.
The proof closely follows the reasoning of the 4-point case.
Calling $N^{(2)}(q)$ the numerator of a term containing the three
denominators $\db{0}\db{1}\db{2}$, and being $j_{max}$ its maximum
rank, one applies \eqnss{eq:4s}{eq:4aaa}
to cast $N^{(2)}(q)$ in a form where all rank $j_{max}$ terms
with three denominators are proportional
to the three numerators given in \eqn{eq:3terms}.
The first two terms are not further reducible, while the iterative use of the
first of Eqs. (\ref{eq:4sa}) reduces the third one
to a sum of contributions proportional to
$[(q+p_0)\cdot \ell_3]^j$ and $[(q+p_0)\cdot \ell_4]^j$ separately, with
$j < j_{max}$.
Then
\bqa
N^{(2)}(q) &=& c(012)+
\sum_{j=1}^{j_{max}} \left\{
 \tld{c}_{1j}(012)[(q+p_0)\cdot \ell_3]^j
+\tld{c}_{2j}(012)[(q+p_0)\cdot \ell_4]^j
        \right\} \nl
&+& \sum_{i=0}^2{\cal O}(\db{i})
+ {\cal O}(\tld{q}^2)\,,
\eqa
where $c(012)$, $\tld{c}_{1j}(012)$ and $\tld{c}_{2j}(012)$
are constants.
Therefore $\tld{c}(q;012)$ must have the form
\bqa \label{eq:6} \tld{c}(q;012)= \sum_{j=1}^{j_{max}} \left\{
 \tld{c}_{1j}(012)[(q+p_0)\cdot \ell_3]^j
+\tld{c}_{2j}(012)[(q+p_0)\cdot \ell_4]^j
        \right\}\,.
\eqa
In the renormalizable gauge, $j_{max}= 3$.
In order to illustrate this fact, a simple argument can be used.
Let us consider the reduction of a $m$-point function of rank $m$.
In the decomposition of Eq.~(\ref{eq:2}),
the 3-point like spurious terms involve $(m-3)$
reconstructed denominators $\db{i}$.
Each one of them can be obtained from a power of $q$ in the numerator $N(q)$
by means for example of Eq.~(\ref{eq:4aaa}).
This leaves at most $m-(m-3)=3$ powers of $q$ available for the
construction of the $\tld{c}(q)$ term.
The same reasoning can be applied to determine $j_{max}=2$ and
$j_{max}=1$ respectively for 2-point and 1-point like spurious terms.

\subsection{The 2-point like spurious terms}
To derive $\tld{b}(q;01)$ it is convenient to rewrite
$q$ in a basis expressed in terms of an auxiliary
arbitrary 4-vector $v$ not parallel to $k_1$
\footnote{\label{foot} We normalize $v$ such that
$(\ell_7 \cdot \ell_8)= -4 (\ell_5 \cdot \ell_6)= - k_1^2$.
Then $\alpha_5 = 2$, $\ell_5 = (k_1+n)/2$ and $\ell_6 = (k_1-n)/4$.}:
\bqa
\label{eq:qb}
q^\mu = -p_0^\mu+ y_1 k_1^\mu+ y_n n^\mu + y_7 \ell_7^\mu + y_8 \ell_8^\mu\,.
\eqa
In a way similar to \eqns{eq:4b}{eq:4c}, $k_1$ and $v$ are decomposed in terms
of two massless 4-vectors $\ell_{5,6}$,
\bqa
k_1   &=& \ell_5 + \alpha_5 \ell_6\,,~~~~~
v  = \ell_6 + \alpha_6 \ell_5\,,
\eqa
and $\ell_{7,8}$ defined as follows
\bqa
\ell_7^\mu &=& <\ell_5| \gamma^\mu | \ell_6]\,,~~
\ell_8^\mu = <\ell_6| \gamma^\mu | \ell_5]\,.
\eqa
Then $n$ is taken to be
\bqa
n   &=& \ell_5 - \alpha_5 \ell_6\,,~~
\eqa
so that it satisfies the two following properties
\bqa
n \cdot k_1 = 0\,,~~{\rm and}~~n^2= -k_1^2\,.
\eqa
Finally, one computes
\bqa
q^\mu =  -p_0^\mu +
       \frac{[(q+p_0) \cdot k_1]}{k_1^2} k_1^\mu
  -    \frac{[(q+p_0) \cdot n]}{k_1^2}   n^\mu
  +    \frac{[(q+p_0) \cdot \ell_8]}{(\ell_7 \cdot \ell_8)}\ell_7^\mu
  +    \frac{[(q+p_0) \cdot \ell_7]}{(\ell_7 \cdot \ell_8)}\ell_8^\mu\,.
\nl
\eqa
By using this basis,
the spurious terms can be determined with the
help of the following

\noindent \underline{Theorems}:
\bqa
\label{eq:7}
 \int d^n \bar q \,
\frac{[(q+p_0)\cdot \ell_7]^j[(q+p_0)\cdot n]^i}{\db{0}\db{1}}&=& 0\,, \nl
 \int d^n \bar q \,
\frac{[(q+p_0)\cdot \ell_8]^j[(q+p_0)\cdot n]^i}{\db{0}\db{1}}&=& 0\,, \nl
 \int d^n \bar q \,
\frac{[(q+p_0)\cdot n]^{2j-1}}{\db{0}\db{1}}&=& 0\,, \nl
 \int d^n \bar q \,
\frac{[(q+p_0)\cdot n]^{2j}
 -r_j \{
[(q+p_0)\cdot k_1]^2-
(q+p_0)^2 k_1^2
      \}^j}{\db{0}\db{1}} &=& 0\,,~~r_1= \frac{1}{3},~~\, r_2= \frac{1}{5}\,,
 \cdots\,, \nl \nl
~~~~\forall j= 1,2,3,\cdots~~~{\rm and}~~~ i= 0,1,2 \cdots
\eqa
Again shifting $q$ and decomposing allows an easy proof.
As for the fourth line of \eqn{eq:7}, both terms in the numerator,
after tensor decomposition, are proportional to $k_1^{2j}$. The factors
$r_j$ are chosen such that their sum is zero.

  Now we are ready to prove that only the terms given
in \eqn{eq:7} contribute to $\tld{b}(q;01)$.
First one rewrites
\bqa
\label{eq:7bis}
q^\mu &=&  G^\mu -\frac{1}{k_1^2}
\left\{
       [(q+p_0) \cdot n]           n^\mu
  +    [(q+p_0) \cdot \ell_7] \ell_8^\mu
  +    [(q+p_0) \cdot \ell_8] \ell_7^\mu
\right\}
 +  \sum_{i=0}^1{\cal O}(\db{i})\,,\nl
G^\mu &\equiv& -p_0^\mu+\frac{k_1^\mu}{k_1^2}
\left[(p_0 \cdot k_1) + \frac{h_1}{2} \right]\,,
\eqa
with $h_1$ given in \eqn{eq:4s}.
Then, calling $N^{(1)}(q)$ the numerator of a term
containing the two
denominators $\db{0}\db{1}$, and being $j_{max}$ its maximum
rank, one replaces each $q$ appearing in $N^{(1)}(q)$
by the r.h.s. of \eqn{eq:7bis}. Therefore, the only possible
generated numerators of degree $j_{max}$ are
\bqa
\label{eq:list}
\begin{tabular}{ll}
          $[(q+p_0) \cdot \ell_7]^j$, &  $[(q+p_0) \cdot \ell_8]^j$, \\
          $[(q+p_0) \cdot \ell_7]^{j_7}[(q+p_0) \cdot n]^{j_n}$, &
          $[(q+p_0) \cdot \ell_8]^{j_8}[(q+p_0) \cdot n]^{j_n}$,  \\
          $[(q+p_0) \cdot n]^j$, &
          $[(q+p_0) \cdot \ell_7]^{i_7} [(q+p_0) \cdot \ell_8]^{i_8}$,
\end{tabular}
\eqa
with $j= j_{max}$, $j_{7,8}+j_n= j_{max}$ and $i_7 + i_8 = j_{max}$.
Due to the first two lines
of \eqn{eq:7}, the first four terms of \eqn{eq:list}
directly give rise to 2-point like spurious terms.
Conversely, the last one can be further reduced by means of the
following identity
\bqa
[(q+p_0) \cdot \ell_7] [(q+p_0) \cdot \ell_8] &=&
4 [(q+p_0) \cdot \ell_5] [(q+p_0) \cdot \ell_6]
-2 (\ell_5 \cdot \ell_6) (q+p_0)^2 \nl
&=& \frac{1}{2}
\left\{
\left((p_0 \cdot k_1)+ \frac{h_1}{2}\right)^2
- k_1^2 m_0^2
-[(q+p_0) \cdot n]^2
\right\}
\nl
&+& \sum_{i=0}^1{\cal O}(\db{i})
+ {\cal O}(\tld{q}^2)\,,
\eqa
where we have used the results of footnote \ref{foot} and
reconstructed denominators as in \eqn{eq:recden}.
Then, the last term of \eqn{eq:list} can be put in the form
\bqa
[(q+p_0) \cdot \ell_7]^{i_7} [(q+p_0) \cdot \ell_8]^{i_8} &=&
\sum_{i=0}^1{\cal O}(\db{i})
+ {\cal O}(\tld{q}^2) \\
&+&
\left\{
\begin{tabular}{rr}
$[(q+p_0) \cdot \ell_8]^{i_8-i_7}
\displaystyle{\sum_{i= 0}^{i_7}}
\delta_i\,
                                   [(q+p_0) \cdot n]^{2i}$,
 & if $i_7 \le i_8$, \\
$[(q+p_0) \cdot \ell_7]^{i_7-i_8}
\displaystyle{\sum_{i= 0}^{i_8}}
\delta_i\,
                                   [(q+p_0) \cdot n]^{2i}$,
 & if $i_8 < i_7$, \\
\end{tabular}
\right. \nonumber
\eqa
where the $\delta_i$ are constants.
All pieces generated by the previous equation
are given by the first five terms of \eqn{eq:list}, but
now with $j \le j_{max}$ and $j_{7,8}+j_n \le j_{max}$. Therefore,
apart from $[(q+p_0) \cdot n]^j$, they are again
taken into account by the first two lines of \eqn{eq:7}; in other words,
they contribute to $\tld{b}(q;01)$.
If $j$ is odd, due to the third line of \eqn{eq:7},
$[(q+p_0) \cdot n]^j$ also gives rise
to spurious 2-point terms. On the contrary,
in order to get the contribution to $\tld{b}(q;01)$ in the case
when $j$ is even,
one has first to subtract powers of
\bqa
[(q+p_0)\cdot k_1]^2-
(q+p_0)^2 k_1^2\,,
\eqa
as performed in the last line of \eqn{eq:7}.
Incidentally, when adding this piece back,
a contribution to $b(01)$ is generated, since
\bqa
[(q+p_0)\cdot k_1]^2-
(q+p_0)^2 k_1^2 =
\left[
\left(
(p_0 \cdot k_1)+ \frac{h_1}{2}
\right)^2
- k_1^2 m_0^2
\right]
+ \sum_{i=0}^1{\cal O}(\db{i})
+ {\cal O}(\tld{q}^2)\,. \nl
\eqa
In summary, no other 2-point like spurious terms are possible
besides those listed in \eqn{eq:7}.

  In the renormalizable gauge there are at most
two powers of $q$ in the numerator. By applying the above
reasonings with $j_{max}= 2$ and $j_{max}= 1$  gives rise to eight
possible spurious $\tld{b}$ terms: \bqa \label{eq:8} \tld{b}(q;01)
&=&
 \tld{b}_{11}(01)[(q+p_0)\cdot \ell_7]
+\tld{b}_{21}(01)[(q+p_0)\cdot \ell_8] \nl
&+&\tld{b}_{12}(01)[(q+p_0)\cdot \ell_7]^2
+\tld{b}_{22}(01)[(q+p_0)\cdot \ell_8]^2 \nl
&+&\tld{b}_{0}(01)[(q+p_0)\cdot n] +\tld{b}_{00}(01)\,K(q;01) \nl
&+& \tld{b}_{01}(01)[(q+p_0)\cdot \ell_7][(q+p_0)\cdot n] \nl
&+& \tld{b}_{02}(01)[(q+p_0)\cdot \ell_8][(q+p_0)\cdot n]\,,
~~~{\rm with}~~~
 \nl \nl
K(q;01) &=& \left\{
[(q+p_0)\cdot n]^2
-\frac{[(q+p_0)\cdot k_1]^2-
       (q+p_0)^2 k_1^2}{3}
\right\}\,.
\eqa

\subsection{The 1-point like spurious terms}
First we decompose
\bqa
\label{eq:qa}
q^\mu = - p_0^\mu + y\, k^\mu+ y_n\, n^\mu + y_7\, \ell_7^\mu + y_8\, \ell_8^\mu\,,
\eqa
where $k$ is an arbitrary 4-vector and $n$, $\ell_7$ and $\ell_8$ are built
up from $k$ and $v$ as in the 2-point case.
Then we make use of the following

\noindent \underline{Theorems}:
\bqa
&&\int d^n \bar q \frac{\prod_{i=1}^{2n-1} (q+p_0) \cdot v_i}{\db{0}} = 0\,, \nl
&&\int d^n \bar q \frac{\prod_{i=1}^{2n}   (q+p_0) \cdot v_i
- r_n (q+p_0)^{2n} g_{\mu_1 \mu_2 \cdots \mu_{2n}}
 v_1^{\mu_1} v_2^{\mu_2} \cdots v_{2n}^{\mu_{2n}}
}{\db{0}} = 0\,,\nl
&&  r_n= (g_{\mu_1 \mu_2 \cdots \mu_{2n}}
     g^{\mu_1 \mu_2} \cdots g^{\mu_{2n-1}\mu_{2n}})^{-1}\,,
~~~\forall n= 1,2,3,\cdots\,,
\eqa
for any 4-vector $v_i$ and where $g_{\mu_1 \mu_2 \cdots \mu_{2n}}$
is the symmetrized product of $n$ metric tensors.

\noindent The proof is just a direct consequence of the fact that
\bqa \int d^n \bar q\, \frac{q^{\mu_1} q^{\mu_2} \cdots
q^{\mu_{2n-1}}}{(\bar q^2- m_0^2)} = 0~~{\rm and}~~ \int d^n \bar
q\, \frac{q^{\mu_1} q^{\mu_2} \cdots q^{\mu_{2n}}}{(\bar q^2-
m_0^2)} \propto  g^{\mu_1 \mu_2 \cdots \mu_{2n}}\,. \eqa
In the
renormalizable gauge at most rank one 1-point functions appear.
Therefore \bqa \label{eq:spua} \tld{a}(q;0) &=& \
 \tld{a}_1(0) [(q+p_0)\cdot k]
+\tld{a}_2(0)[(q+p_0)\cdot n] \nl
&+&\tld{a}_3(0)[(q+p_0)\cdot \ell_7]
+\tld{a}_4(0)[(q+p_0)\cdot \ell_8]\,.
\eqa

\subsection{The 0-point like spurious term}
As already pointed out, $\tld{P}(q)= 0$ in the renormalizable gauge.
We can prove this statement simply by counting the powers of $q$ in
\eqn{eq:2}. The last term on the r.h.s. contains
$2 m$ powers of $q$ in the $m$ reconstructed denominators $\db{i}$.
Since $N(q)$ on the l.h.s. is at most of rank $m$ and the other
terms on the r.h.s. contain at most $2 m - 1$ powers of $q$,
in order to satisfy \eqn{eq:2} we should have $\tld{P}(q)= 0$.

In more general gauges, $\tld{P}(q)$ gives a contribution, polynomial in $q$,
to the integrand of the amplitude.
Terms like that vanish, upon integration, in dimensional regularization.
 This is why we classified $\tld{P}(q)$ among the spurious term.
Off course $\tld{P}(q)$ is never needed, in any gauge, {\em if}
all the other coefficients in \eqn{eq:2}
can be determined without making use of its actual form.
This is the case, as we shall see in the next Section.
However, if necessary, after all the other coefficients are known,
 $\tld{P}(q)$ is easily computed as the difference between $N(q)$
and the first four lines of the r.h.s. of \eqn{eq:2},
divided by all the denominators.

\section{\label{sec:4}{Extracting the coefficients of the scalar loop functions}}
Being interested here in the coefficients of the scalar loop
functions, we can set everywhere \bqa \tld{q}^2= 0\,, \eqa so that,
in particular \bqa \label{eq:qtld0} \db{i} \to  \d{i} \equiv (q +
p_i)^2-m_i^2\,. \eqa The ``error'' induced by the above replacement
is at the level of the rational part of the amplitude, as we shall
see in Section \ref{sec:5}, where we will also use the fact that the
$\tld{q}^2$ terms are always connected to the masses in the
denominators to reconstruct the information we are missing with the
replacement of \eqn{eq:qtld0}.

  Our approach to the problem is suggested by the structure
of \eqn{eq:2} itself. Choosing particular values of $q$ such that
4, 3, 2 or 1 denominators vanish allows one to reduce the problem
to the solution of simpler sub-systems of equations.
  To illustrate how this works we concentrate again on the particular
choice $i_0= 0$, $i_1= 1$, $i_2= 2$ and $i_3= 3$ and derive
explicit formulae for the coefficients
${d}(0123)$, ${c}(012)$, ${b}(01)$ and ${a}(0)$
as well as for the coefficients of the corresponding spurious terms
\footnote{For any other choice of indices the procedure is the same.}.
  As we shall see, the latter information is also needed when
iterating the algorithm: in order to solve
for the $c(012)$ coefficient, one has to know the coefficients of all
terms with 4 denominators, including the spurious ones.
In order to solve for $b(01)$, one needs, in addition, all
the terms with 3 denominators and so on.
\subsection{The coefficient of the 4-point functions}
We look for a $q$ such that \bqa \label{eq:8a} \d{0}= \d{1}= \d{2}=
\d{3} = 0\,. \eqa
By writing the 4-vector $q$ as \bqa \label{eq:8aa} q^\mu = -p_0^\mu
+\sum_{i=1}^{4}  x_i\,\ell_i^\mu\,, \eqa with $\ell_i$ given in
\eqnss{eq:4b}{eq:4d}, one obtains a system of equations for the
$x_i$: \bqa \label{eq:9} 0&=& \gamma(x_1 x_2 -4 x_3 x_4)-d_0 \nl
0&=& d_0-d_1+\gamma(x_1\alpha_1+ x_2) \nl 0&=&
d_0-d_2+\gamma(x_2\alpha_2+ x_1) \nl 0&=& d_0 -d_3 +2 \left[
 x_1 (k_3 \cdot \ell_1)
+x_2 (k_3 \cdot \ell_2)
+x_3 (k_3 \cdot \ell_3)
+x_4 (k_3 \cdot \ell_4)
               \right]
\,,
\eqa
where $k_i$ is given in \eqn{eq:4bb} and
\bqa
\label{eq:10}
d_i \equiv m_i^2-k_i^2\,.
\eqa
\noindent There are two possible solutions
\bqa
\label{eq:solut4}
({q_0}^{\pm})^\mu= -p_0^\mu+
 x_1^0 \ell_1^\mu
+x_2^0 \ell_2^\mu
+{x_3}^\pm \ell_3^\mu
+{x_4}^\pm \ell_4^\mu\,,
\eqa
with
\bqa
\label{eq:sold1}
&&x_1^0 = \frac{\beta}{\gamma}
[d_2 - \alpha_2 d_1 -d_0 (1- \alpha_2)]\,, \nl
&&x_2^0 = \frac{\beta}{\gamma}
[d_1 - \alpha_1 d_2 -d_0 (1- \alpha_1)]\,, \nl
&& A\, {x_3^\pm}^2 +B\, x_3^\pm - C = 0\,, \nl
&&x_4^\pm = \frac{C}{x_3\pm}\,,
\eqa
and
\bqa
\label{eq:sold2}
A &=& -\frac{(k_3 \cdot \ell_3)}{(k_3 \cdot \ell_4)}\,,~~
B = \frac{d_3-d_0
-2 x_1^0 (k_3 \cdot \ell_1)
-2 x_2^0 (k_3 \cdot \ell_2)
}{2\,(k_3 \cdot \ell_4)} \,,\nl
C &=& \frac{1}{4} \left(x_1^0 x_2^0-\frac{d_0}{\gamma}
                  \right)\,.
\eqa Notice that we {\em need} two solutions to be able to determine
both ${d}(0123)$ and $\tld{d}(0123)$, and that the existence of more
than one solution is a consequence of the quadratic nature of the
system in \eqn{eq:9}. By putting both $q_0^\pm$ in \eqn{eq:2}, and
recalling the form of $\tld{d}(q;0123)$ given in \eqn{eq:3a}, one
finds
\bqa
N(q_0^\pm) &=& [d(0123) + \tld{d}(0123)\, T(q_0^\pm)] \prod_{i\ne 0,1,2,3}
\d{i} (q_0^\pm)
\,.
\eqa
Then, by defining
\bqa
R(q_0^\pm) \equiv \frac{N(q_0^\pm)}{\prod_{i\ne 0,1,2,3}
\d{i} (q_0^\pm)
}\,,
\eqa
it is possible to extract $d$ and $\tld{d}$
\bqa
   d(0123)
&=& \frac{R(q_0^-)T(q_0^+) - R(q_0^+)T(q_0^-)}{T(q_0^+)-T(q_0^-)}\,, \nl
\tld{d}(0123)
&=& \frac{R(q_0^+)-R(q_0^-)}{T(q_0^+)-T(q_0^-)}\,.
\eqa
Notice that, in terms of $x_{3,4}^\pm$, one rewrites
\bqa
T(q_0^\pm) &=& 2 \gamma \left[
 x_3^\pm (k_3 \cdot \ell_3)
-x_4^\pm (k_3 \cdot \ell_4)
\right]\,, \nl
T(q_0^+) &=&-T(q_0^-)\,.
\eqa
Then
\bqa
   d(0123)   &=&  \frac{1}{2}[R(q_0^+)+ R(q_0^-)]\,, \nl
\tld{d}(0123)&=&  \frac{1}{2}\frac{R(q_0^+)- R(q_0^-)}{T(q^+)}\,.
\eqa
The two above equations {\em do not depend} on the rank of
the tensors in the amplitude. When $N(q)= 1$, they allow
a trivial decomposition of any $m$-point scalar loop function with
$m > 4$ to boxes, as we shall see in Section \ref{appli}.
\subsection{The coefficient of the 3-point functions}
At this stage all $d$ and $\tld{d}$ coefficients are known.
When $q$ is such that
\bqa
\label{eq:8b}
\d{0}= \d{1}= \d{2}= 0~~~{\rm and}~~~\d{i} \ne 0~~ \forall i \ne 0,1,2
\eqa
\eqn{eq:2} reads
\bqa
N(q) &-& \sum_{2 <i_3}[d(012i_3)
           + \tld{d}(q;012i_3)]\prod_{i \ne 0,1,2,i_3}\d{i}(q) \nl\nl
&\equiv& R^\prime(q) \prod_{i \ne 0,1,2} \d{i}(q)
= [c(012) + \tld{c}(q;012)]\prod_{i \ne 0,1,2} \d{i}(q)\,,
\eqa
and one can extract $c(012)$, together with all the six
$\tld{c}_{ij}(012)$  coefficients of \eqn{eq:6},
by computing $R^\prime(q)$ at seven different  $q$'s that fulfill \eqn{eq:8b}.
For a $q$ written as in \eqn{eq:8aa} that happens when
\bqa
\label{eq:11}
x_1 &=& x_1^0 \nl
x_2 &=& x_2^0 \nl
x_3 x_4 &=& C\,,
\eqa
where $x^0_{1,2}$ and $C$ are given in \eqns{eq:sold1}{eq:sold2}.
There is now an infinite number of solutions, which we parametrize
by imposing the extra condition
\bqa
(q+p_0)\cdot \ell_3 = \pm \sqrt{C} e^{i \pi /k} (\ell_3 \cdot \ell_4)\,,
~~k = 1,2,3,\cdots\,.
\eqa
Then
\bqa
x_4 = \pm \sqrt{C} e^{i \pi /k} \equiv x_{4k}^\pm\,,~~
x_3 = \frac{C}{x_{4k}^\pm} = \pm \sqrt{C} e^{-i \pi /k}
 \equiv x_{3k}^\pm\,,
\eqa
and the complete solution reads
\bqa
\label{eq:solc}
(q^\pm_k)^\mu = -p_0^\mu +
 x_1^0\ell_1^\mu
+x_2^0\ell_2^\mu +x_{3k}^\pm \ell_3^\mu, +x_{4k}^\pm \ell_4^\mu\,.
\eqa Finally, with the help of the relation \bqa [(q^\pm_k+p_0)\cdot
\ell_4] [(q^\pm_k+p_0)\cdot \ell_3]= C (\ell_3 \cdot \ell_4)^2 \,,
\eqa one writes \bqa \label{eq:12} R^\prime(q^\pm_k)= \sum_{j=
-3}^{3} c_j [\pm e^{i \pi /k}]^j\,, \eqa with \bqa \tld{c}_{1j}(012)
~=~ \frac{c_j   }{C^{j/2}\,(\ell_3 \cdot \ell_4)^j}\,,~~~~
\tld{c}_{2j}(012) ~=~ \frac{c_{-j}}{C^{j/2}\,(\ell_3 \cdot
\ell_4)^j}~~~ {\rm and}~~~c(012) ~=~ c_0\,. \eqa In \app{appA}, we
explicitly determine all $c_j$'s of \eqn{eq:12} by choosing 7
different solutions of the form given in \eqn{eq:solc}.

\subsection{The coefficient of the 2-point functions}
At this stage all $d$, $\tld{d}$, $c$ and $\tld{c}$  coefficients are known.
When $q$ is such that
\bqa
\label{eq:14}
\d{0}= \d{1}= 0~~~{\rm and}~~~\d{i} \ne 0~~ \forall i \ne 0,1\,
\eqa
\eqn{eq:2} reads
\bqa
 N(q)
&-& \sum_{1 < i_2 < i_3}[d(01i_2i_3)
+ \tld{d}(q;01i_2i_3)]  \prod_{i \ne 0, 1,i_2,i_3}\d{i} \nl \nl
&-& \sum_{1 < i_2}[c(01i_2)
+ \tld{c}(q;01i_2)] \prod_{i \ne 0, 1, i_2}\d{i} \nl \nl
&\equiv& {R}^{\prime\prime}(q) \prod_{i\ne 0,1}\d{i}(q)
 = [b(01) + \tld{b}(q;01)]\prod_{i\ne 0,1}\d{i}(q)\,,
\eqa
and one can extract $b(01)$
together with all the eight
$\tld{b}(01)$  coefficients of \eqn{eq:8},
by computing ${R}^{\prime\prime}(q)$
at nine different  $q$'s that fulfill \eqn{eq:14}.
That happens when, for
a $q$ written as in \eqn{eq:qb},
\bqa
\label{eq:15}
y_1   &=& \frac{d_1-d_0}{2 k_1^2} \equiv y_1^0 \,,\nl
y_n^2 &=& (y_1^0)^2-\frac{d_0+2 k_1^2 y_7 y_8}{k_1^2}\,,
\eqa
where we used our normalization condition $(\ell_7 \cdot \ell_8) = - k_1^2$.
We impose now, as an extra requirement,
that $K(q;01)$ defined in \eqn{eq:8} vanishes . This implies
\bqa
y_1 &=& y_1^0 \,,\nl
y_n &=& \pm\sqrt{\frac{1}{3}\left({(y^0_1)}^2-\frac{d_0}{k_1^2}\right)}
\equiv \pm \sqrt{F}\,, \nl
y_7 &=& \frac{F}{y_8}\,.
\eqa
We fix the remaining freedom by imposing
\bqa
(q+p_0) \cdot \ell_7 =
\pm \sqrt{F} e^{i\pi/k} (\ell_7 \cdot \ell_8)\,,
\eqa
that implies
\bqa
y_8 &=& \pm \sqrt{F} e^{i\pi/k} \equiv y^\pm_{8k} \nl
y_7 &=& \frac{F}{y^\pm_{8k}}=
\pm \sqrt{F} e^{-i\pi/k} \equiv y^\pm_{7k}\,.
\eqa
Then, this class of solutions can be parametrized as follows
\bqa
\label{eq:solb}
(q^\pm_{lk})^\mu = - p_0^\mu+y_1^0\, k_1^\mu + l \sqrt{F}\, n^\mu
+y^\pm_{7k}\, \ell_7^\mu
+y^\pm_{8k}\, \ell_8^\mu\,,~~~~{\rm with} ~~l = \pm 1\,.
\eqa
Since
\bqa
(q+p_0) \cdot \ell_8 = \frac{F (\ell_7 \cdot \ell_8)^2}{(q+p_0)\cdot \ell_7}\,,
\eqa
we can rewrite (see \eqn{eq:8})
\bqa
R^{\prime \prime}(q^\pm_{lk})= l \sum_{j= -1}^{1}
\beta_j\, [\pm e^{i\pi/k}]^j
+ \sum_{j= -2}^{2} b_j\,[\pm e^{i\pi/k}]^j\,,
\eqa
where
\bqa
\begin{tabular}{lll}
$b(01) = b_0$  & $\tld{b}_{1j}(01) =
\displaystyle{\frac{b_j}{F^{j/2}\,(\ell_7 \cdot \ell_8)^j}}$ &
 $\tld{b}_{2j}(01) =
\displaystyle{\frac{b_{-j}}{F^{j/2}\,(\ell_7 \cdot \ell_8)^j}}$
\\\\
$\tld{b}_{01}(01) =
\displaystyle{
\frac{\beta_1}{F (\ell_7 \cdot \ell_8)^2}}$ &
$\tld{b}_{02}(01) =
\displaystyle{
\frac{\beta_{-1}}{F (\ell_7 \cdot \ell_8)^2}}$ &
$\tld{b}_0(01) =
\displaystyle{
\frac{\beta_0}{F^{1/2} (\ell_7 \cdot \ell_8)}}\,.$
\end{tabular}
\eqa To determine the last coefficient $\tld{b}_{00}(01)$ we need to
introduce a different solution, for which $K(q;01)$ does not vanish.
We call $q_0$ such a solution and choose it satisfying the condition
$y_7 = y_8 =0$. Then, by insertion in \eqns{eq:15}{eq:8}, one finds
\bqa
q_0^\mu = -p_0^\mu +y_1^0\, k_1^\mu + \sqrt{3F}\, n^\mu\,,
\eqa
and
\bqa \label{eq:bq0} R^{\prime \prime}(q_0) &=& b(01)  +
\tld{b}_0(01) \,\,[(q_0+p_0) \cdot n] + \tld{b}_{00}(01)\,\,
K(q_0;01)\nl &=&
 b(01) -\tld{b}_0(01) \,\,\sqrt{3F} k_1^2
+2\tld{b}_{00}(01)\,\, k_1^4 F\,. \eqa
The complete solution for all the
coefficients is given in \app{appB}.

\subsection{The coefficient of the 1-point functions}
In massless theories all 1-point functions, namely all tadpoles,
vanish, also implying that, in such cases, one does not need to know
all the $\tld{b}$ coefficients: only the coefficients of the scalar
2-point functions are needed, each of them can be determined in
terms of just four solutions of \eqn{eq:14} (see \app{appB}).
However, in general, also the coefficients of the tadpoles are
required. Therefore we show how to extract them.

At this stage we assume to know all the
$d$, $\tld{d}$, $c$, $\tld{c}$, $b$ and $\tld{b}$ coefficients and,
when $q$ is such that
\bqa
\label{eq:sola}
\d{0}= 0~~~{\rm and}~~~\d{i} \ne 0~~ \forall i \ne 0\,,
\eqa
\eqn{eq:2} reads
\bqa
 N(q)
&-& \sum_{0 <i_1 < i_2 < i_3}[d(0i_1i_2i_3)
+ \tld{d}(q;0i_1i_2i_3)]  \prod_{i \ne 0, i_1, i_2,i_3}\d{i} \nl \nl
&-& \sum_{0 <i_1 < i_2}[c(0i_1i_2)
+ \tld{c}(q;0i_1i_2)] \prod_{i \ne 0, i_1, i_2}\d{i} \nl \nl
&-& \sum_{0< i_1}[b(0i_1)
+ \tld{b}(q;0i_1)] \prod_{i \ne 0, i_1}\d{i} \nl \nl
&\equiv& {R}^{\prime\prime\prime}(q) \prod_{i\ne 0}\d{i}(q)
 = [a(0) + \tld{a}(q;0)]\prod_{i\ne 0}\d{i}(q)\,.
\eqa
Then, we parametrize $q$ as in  \eqn{eq:qa} and choose two solutions
of \eqn{eq:sola} such that $y_n = y_7= y_8 =0$:
\bqa
(q_0^{\pm})^\mu =  -p_0^\mu \pm \sqrt{\frac{d_0}{k^2}}\,k^\mu\,.
\eqa
Therefore, by recalling \eqn{eq:spua}, we can write \bqa R^{\prime
\prime \prime}(q_0^+) &=& a(0) + \tld{a}_1(0)
  \sqrt{\frac{d_0}{k^2}} k^2 \nl
R^{\prime \prime \prime}(q_0^-) &=& a(0) - \tld{a}_1(0)
  \sqrt{\frac{d_0}{k^2}} k^2\,,
\eqa
so that
\bqa
a(0) = \frac{R^{\prime \prime \prime}(q_0^+)
+R^{\prime \prime \prime}(q_0^-)}{2}\,.
\eqa
Notice that $\tld{a}(q;0)$ is never needed,
because it would be necessary only to extract $\tld{P}(q)$,
that, as already observed, is irrelevant.

\section{\label{sec:5} Reconstructing the rational part of the amplitude}
Until now we have assumed $\tld{q}^2= 0$. As already discussed, this
is enough to reconstruct the coefficients of the 4-3-2-1-point loop
functions, but rational parts are missing. In the renormalizable
gauge the only possible contributions to those rational terms come
from the following extra scalar integrals introduced in
\cite{delAguila:2004nf} \footnote{The powers of $\tld{q}^2$ are
dictated by the maximum rank of the loop tensors in $A(\bar q)$.}
\bqa \label{eq:rat} \int d^n \bar{q}
\frac{\tld{q}^4}{\db{i}\db{j}\db{k}\db{l}} &=& - \frac{i \pi^2}{6} +
\cal{O}(\epsilon)\,,\nl \int d^n \bar{q}
\frac{\tld{q}^2}{\db{i}\db{j}\db{k}}       &=& - \frac{i \pi^2}{2} +
\cal{O}(\epsilon)\,,\nl \int d^n \bar{q}
\frac{\tld{q}^2}{\db{i}\db{j}}             &=& - \frac{i \pi^2}{2}
\left[m_i^2+m_j^2-\frac{(p_i-p_j)^2}{3} \right]   +
\cal{O}(\epsilon)\,. \eqa We checked that they reproduce the
rational terms listed in \cite{Xiao:2006vr}. Therefore, in our
language, the coefficients of the integrals in \eqn{eq:rat} are just
the coefficients of the maximum powers of $\tld{q}^2$ contained in
the $d(ijkl)$, $c(ijk)$ and $b(ij)$ once $\tld{q}^2$ is reintroduced
through the mass shift \bqa m_i^2 \to m_i^2 -\tld{q}^2. \eqa With
the above replacement the coefficients get a dependence on
$\tld{q}^2$ and one can expand: \bqa \label{eq:ratexp}
d(ijkl;\tld{q}^2) &=&  d(ijkl)
                     +\tld{q}^2 d^{(2)}(ijkl)
                     +\tld{q}^4 d^{(4)}(ijkl)\,, \nl
c(ijk;\tld{q}^2) &=&   c(ijk)
                     +\tld{q}^2 c^{(2)}(ijk)\,, \nl
b(ij;\tld{q}^2) &=&   b(ij)
                     +\tld{q}^2 b^{(2)}(ij) \,.
\eqa $d^{(4)}(ijkl)$, $c^{(2)}(ijk)$ and $b^{(2)}(ij)$ are then the
coefficients of the first, second and third integral of
\eqn{eq:rat}, respectively. They can be either computed numerically
\bqa d^{(4)}(ijkl) &=&  \lim_{\tld{q}^2 \to \infty}
\frac{d(ijkl;\tld{q}^2)}{\tld{q}^4}
 \,, \nl
c^{(2)}(ijk)  &=&  \lim_{\tld{q}^2 \to \infty}
\frac{c(ijk;\tld{q}^2)}{\tld{q}^2} \,, \nl b^{(2)}(ij)   &=&
\lim_{\tld{q}^2 \to \infty} \frac{b(ij;\tld{q}^2)}{\tld{q}^2} \,,
\eqa
or as solutions of systems obtained by evaluating \eqn{eq:ratexp} at
different $\tld{q}^2$. For instance: \bqa d^{(4)}(ijkl)&=&
\frac{d(ijkl;1)+d(ijkl;-1)-2 d(ijkl)}{2}\,, \nl c^{(2)}(ijk) &=&
c(ijk;1)- c(ijk)\,, \nl b^{(2)}(ij)  &=& b(ij;1)- b(ij)\,. \eqa A
small example of the described approach is given in the next
Section.

\section{\label{appli}Applications and tests}
We tested the whole method on the reduction of a rank four 4-point
tensor integral \bqa \label{eq:ten4} \int d^n \bar{q} \frac{q^\mu
q^\nu q^\rho q^\sigma}{\db{0}\db{1}\db{2}\db{3}}\, \eqa to scalar
functions. We have been able to correctly extract the coefficients
of the scalar integrals. In addition, when reducing the tensor in
\eqn{eq:ten4} with the techniques of \cite{delAguila:2004nf}, we
have been able to also test the coefficients of the spurious terms.

As for the rational terms, we explicitly show here, as an
illustrative example, the extraction of the coefficient of
\bqa \int d^n \bar{q} \frac{\tld{q}^2}{\db{0}\db{1}\db{2}}
\eqa
from
\bqa \int d^n \bar{q} \frac{q^\mu q^\nu}{\db{0}\db{1}\db{2}}\,,
\eqa
where, for simplicity, we have put $p_0= 0$. Then, in this case
\bqa
N(q) = q^\mu q^\nu = R^\prime(q)\,. \eqa
We first observe that, in
the limit $\tld{q}^2 \to \infty$, one can use the following
asymptotic form of the solutions given in \eqn{eq:solc}
\bqa (q^\pm_1)^\mu &=& \mp \sqrt{C_{\infty}}\, (\ell_3^\mu +
\ell_4^\mu)\,,~~~ (q^\pm_2)^\mu  = \mp i\sqrt{C_{\infty}}\,
(\ell_3^\mu - \ell_4^\mu)\nl ~~~{\rm with}~~~ C_{\infty} &\equiv&
\lim_{\tld{q}^2 \to \infty} C = \frac{\tld{q}^2 }{4 \gamma}\,.
\eqa
Therefore, one obtains, for the coefficient  $c_0$ in \eqn{eq:sysc}
\bqa \lim_{\tld{q}^2 \to \infty} \frac{c_0}{\tld{q}^2} =
\frac{\ell_3^\mu \ell_4^\nu +\ell_4^\mu \ell_3^\nu}{4\gamma}\,, \eqa
Then
\bqa
c^{(2)}(012)= \frac{\ell_3^\mu \ell_4^\nu +\ell_4^\mu
\ell_3^\nu}{4\gamma}\,,
\eqa
in agreement with the result obtained
in \cite{delAguila:2004nf}.

Another rather straightforward application of the method is the
reduction of the  scalar $n$-point functions, with $n\ge 5$, in
terms of box functions. It will allow the reader to follow the
reasoning of the reduction in a simple case. It should be mentioned
that the content of this derivation, amazingly enough, goes back to
the year 1965, in the work of Melrose~\cite{Melrose:1965kb} and
K\"all\'en and Toll~\cite{kallen:1965}.

\noindent In the conventional approach one can easily prove
~\cite{Denner:1991kt,Melrose:1965kb} the following
relation
\begin{equation}
\left|
\begin{array}{ccccc}
  I^N    & -I^{N-1}(0)    & -I^{N-1}(1)    & \ldots & -I^{N-1}(N-1) \\
  1      & Y_{0\,0}   & Y_{0\,1}   & \ldots & Y_{0\,N-1} \\
  1      & Y_{1\,0}   & Y_{1\,1}   & \ldots & Y_{1\,N-1} \\
  \vdots & \vdots     & \vdots     & \vdots & \vdots \\
  1      & Y_{N-1\,0} & Y_{N-1\,1} & \ldots & Y_{N-1\,N-1}
\end{array}
\right | =0\,,
\label{det=0}
\end{equation}
where $I^N$ is the $N$-point scalar function
\bqa
I^N = \int d^n \bar q \frac{1}{\db{0} \cdots \db{N-1}}\,,
\eqa
$I^{N-1}(i)$ is the
$N-1$-point function with the $i$-th propagator missing  and
\begin{equation}
Y_{ij}=m_i^2+m_j^2-(p_i-p_j)^2, \,\,\, i=0,\ldots,N \,\,\,
j=0,\ldots,N\,.
\end{equation}
By repeated use of \eqn{det=0} we may express the $N$-point function
in terms of $4$-point functions with coefficients expressible in
terms of the determinants of $Y$ matrices. For instance
\begin{equation}
N=5: \,\,\,\,\,\, I^5 =-\sum_{i=0}^{4}
\frac{det_i(Y^{(5)})}{det(Y^{(5)})} I^4(i)\,,
\end{equation}
where $Y^{(5)}$ is the $5\times5$ $Y$ matrix, and $det_i$ represents
the determinant of matrix $Y$ where all elements of the $i$-th
column have been replaced by 1.

\noindent Similarly we obtain

\begin{equation} N=6: \,\,\,\,\,\, I^6
=-\sum_{i=0}^{5} \frac{det_i(Y^{(6)})}{det(Y^{(6)})} I^5(i)\,.
\end{equation}

Let us now see how these formula get simplified using the method
described so far. For the 5-point function we get

\begin{equation}
I^5 =\sum_{i=0}^{4} d_i I^4(i)
\end{equation}

with
\begin{equation}
d_i=\frac{1}{2}\left(\frac{1}{D_i(q^+_{(i)})}+\frac{1}{D_i(q^-_{(i)})}\right)\,,
\end{equation}
whereas the 6-point function reads
\begin{equation}
I^6 =\sum_{i<j}\sum_{i,j=0}^{5} d_{ij} I^4(ij)\,,
\end{equation}
where $I^4(ij)$ is obtained from $I^6$ by dropping the propagators
$\db{i}$ and $\db{j}$ and where
\begin{equation}
d_{ij}=\frac{1}{2}\left(\frac{1}{D_i(q^+_{(ij)})D_j(q^+_{(ij)})}
+\frac{1}{D_i(q^-_{(ij)})D_j(q^-_{(ij)})}\right)\,.
\end{equation}
In the above equations $q^\pm_{(i)}$ are the two solutions
given in \eqn{eq:solut4} when all the propagators, except $\d{i}$,
are zero. Analogously  $q^\pm_{(ij)}$ are the solutions
when $\d{i}$ and $\d{j}$ are the only non vanishing propagators.

It is quite straightforward to prove that

\begin{equation}
\frac{1}{2}\left(\frac{1}{D_i(q^+_{(i)})}+\frac{1}{D_i(q^-_{(i)})}\right)=
-\frac{det_i(Y^{(5)})}{det(Y^{(5)})}\,.\label{5ps}
\end{equation}
This is because the Gram determinant of $q,p_1,p_2,p_3,p_4,p_5$
should be zero, which results to

\small
\bqa
\left|
\begin{array}{ccccc}
  2 D_0+Y_{00}         & D_1-D_0+Y_{10}-Y_{00}        & D_2-D_0+Y_{20}-Y_{00}        & \ldots & D_5-D_0+Y_{20}-Y_{00}       \\
  D_1-D_0+Y_{10}-Y_{00} & Y_{11}-Y_{10}-Y_{01}+Y_{00}   & Y_{12}-Y_{10}-Y_{02}+Y_{00}   & \ldots & Y_{15}-Y_{10}-Y_{05}+Y_{00} \\
  D_2-D_0+Y_{20}-Y_{00} & Y_{21}-Y_{20}-Y_{01}+Y_{00}   & Y_{22}-Y_{20}-Y_{02}+Y_{00}   & \ldots & Y_{25}-Y_{20}-Y_{05}+Y_{00} \\
  \vdots               & \vdots                       & \vdots                       & \vdots & \vdots                     \\
  D_5-D_0+Y_{20}-Y_{00} & Y_{51}-Y_{50}-Y_{01}+Y_{00}   & Y_{52}-Y_{50}-Y_{02}+Y_{00}   & \ldots & Y_{55}-Y_{50}-Y_{05}+Y_{00}
\end{array}
\right | =0
 \label{gramd}  \nl
\eqa

\normalsize
\noindent when on takes into account that

\[ 2 q^2=2 D_0 +2 m_0^2\,, \]
\[ 2 q\cdot p_j=D_j-D_0+Y_{0j}-Y_{00}\,, \]
and
\[ 2p_i\cdot p_j =Y_{ij}-Y_{i0}-Y_{0j}+Y_{00}\,. \]

\noindent Taking \eqn{gramd} at the point $q=q_{(i)}$ we end up with a second
order equation for $D_i$ given by
\[a D_i^2 +b D_i+c=0\,,\]
with $b=-2 det_i(Y^{(5)})$ and $c=det(Y^{(5)})$. An analytical proof
for arbitrary $N$, can be found in~\cite{Melrose:1965kb}.

\section{Conclusions}
We have shown how computing the {\em integrand} of any one-loop
amplitude at special values of the integration momentum allows the
one-shot reconstruction of all the coefficients of the scalar loop
functions and of the rational terms.
 Then, by simply multiplying those coefficients by the
known scalar integrals, the computation of the amplitude becomes
trivial.
 Our method should be  particularly useful in the case when
recursive techniques are used to numerically compute the integrand.
 We plan to investigate this subject in the near future.

\section*{Acknowledgments}
We thank Pierpaolo Mastrolia and Andre van Hameren for useful
discussions. RP and GO acknowledge the financial support of the ToK
Program ``ALGOTOOLS'' (MTKD-CD-2004-014319). The research of RP was
also supported in part by MIUR under contract 2004021808\_009.

\section*{Appendices}
\appendix
\section{\label{appA}The system for the coefficients of the 3-point
functions}

We choose the following seven solutions (see \eqn{eq:solc}) \bqa
q^\pm_1\,,~~q^\pm_2\,,~~q^\pm_3\,~~{\rm and}~~q^+_6\,, \eqa and
define the combinations \bqa T^\pm(q_k) &\equiv&
\frac{R^\prime(q^+_k)\pm R^\prime(q^-_k)}{2} \,. \eqa Then, from
\eqn{eq:12}, one obtains, for the even powers of $j$, the system
\bqa \label{eq:evenc} \left\{
\begin{tabular}{llll}
$T^+(q_1)$ =& $+c_{-2}            $ & $+ c_0       $ & $ + c_2$      \\
$T^+(q_2)$ =& $-c_{-2}            $ & $+ c_0       $ & $ - c_2$      \\
$T^+(q_3)$ =& $+c_{-2}e^{-2i\pi/3}$ & $+ c_0       $ & $ +c_2 e^{2i\pi/3}$
\end{tabular} \right.\,,
\eqa
whose solution is
\bqa
\label{eq:sysc}
 c_0       &=& \frac{T^+(q_1)+T^+(q_2)}{2}\,, \nl
 c_{\pm 2} &=& \left[
 \frac{T^+(q_1)-T^+(q_2)}{2}
 -e^{\pm 2 i \pi/3}(T^+(q_3)-c_0)
\right]
\frac{1}{1-e^{\mp 2 i \pi/3}}\,.
\eqa
For the odd powers of $j$ one gets instead
\bqa
\left\{
\begin{tabular}{lllll}
$T^-(q_3)$ =& $-  c_{-3}            $ & $ + c_{-1}e^{-i\pi/3} $ & $ +c_{1}e^{i\pi/3}$  & $-c_{3}$ \\
$T^-(q_2)$ =& $+i  c_{-3}           $ & $-i c_{-1} $ & $ +i c_{1}$& $-ic_{3}$ \\
$T^-(q_1)$ =& $-  c_{-3}$ & $-  c_{-1} $ & $ -c_{1}$  & $-c_{3}$ \\
$T^0(q_6)$ =& $-i c_{-3}$ & $+ c_{-1}e^{-i\pi/6}  $ & $ + c_{1}e^{i\pi/6}$ & $+ic_{3}$
\end{tabular}
\right. \,,
\eqa
where
\bqa
 \nl
T^0(q_k)   &\equiv& R^\prime(q^+_k)-\sum_{j= -1}^{1} c_{2j} [e^{i \pi /k}]^{2j}\,
\eqa
is known because the coefficients $c_{0,\pm2}$ have already been determined.
The solution reads
\bqa
c_{\pm1 } &=&
\frac{
  [T^-(q_3)-T^-(q_1)]        (1+e^{\pm i\pi/3})
\mp i[T^0(q_6)+T^-(q_2)]     (1+e^{\mp i\pi/3})
}{3}\,, \nl
c_{\pm3}  &=& -c_{\mp1}-\frac{T^-(q_1)\mp i T^-(q_2)}{2}\,.
\eqa

\section{\label{appB}The system for the coefficients of the 2-point
functions}

We choose the following nine solutions  (see \eqn{eq:solb}) \bqa
&&q^+_{\pm1 1}\,,~~q^+_{\pm1 2}\,,~~q^+_{\pm1 3}\,,~~
  q^-_{1-1}\,~~ q^-_{1-2}\,,~~ q_0\,,
\eqa and build up the combinations \bqa S^\pm(q^\pm_k) \equiv
\frac{R^{\prime \prime}(q^\pm_{+1k})
                        \pm R^{\prime \prime}(q^\pm_{-1k})}{2}\,.
\eqa
Therefore the $\beta$ coefficients satisfy the system
\bqa
\left\{
\begin{tabular}{llll}
$S^-(q^+_1)$ =& $-\beta_{-1}              $ & $+ \beta_0  $ & $ - \beta_1$  \\
$S^-(q^+_2)$ =& $- i\beta_{-1}            $ & $+ \beta_0  $ & $ +i \beta_1$ \\
$S^-(q^+_3)$ =& $\beta_{-1}e^{-i\pi/3}$
& $+ \beta_0  $& $ \beta_1 e^{i\pi/3}$
\end{tabular} \right.\,,
\eqa
whose solution reads
\bqa
\beta_{\pm1} &=& \frac{(1+ e^{\mp i\pi/3})
[S^-(q^+_2)-S^-(q^+_1)]-(1 \mp i) [S^-(q^+_3)-S^-(q^+_1)]}{\pm i(3-\sqrt{3})}\,, \nl
\beta_0 &=& S^-(q^+_1) +(\beta_{-1}+\beta_1)\,.
\eqa
Next, by defining
\bqa
T^\pm(q_k) \equiv \frac{S^+(q^+_{k})
                    \pm S^+(q^-_{k})}{2}\,,
\eqa one finds \bqa \left\{
\begin{tabular}{lll}
$T^-(q_1)$ =& $-   b_{-1}            $ & $ - b_1$  \\
$T^-(q_2)$ =& $- i b_{-1}            $ & $ +i b_1$
\end{tabular} \right.\,,
\eqa
whose solution is
\bqa
b_{\pm 1} = -\frac{1}{2} \left[
T^-(q_1)\pm i T^-(q_2)
\right]\,.
\eqa
Now that $\beta_0$, $\beta_{-1}$, $\beta_1$, $b_{-1}$ and  $b_{1}$ are known
we define
\bqa
T^0(q_3) \equiv  R^{\prime \prime}(q^+_{+13})-\sum_{j= -1}^{1}\beta_j
\, [e^{i\pi/3}]^j
- b_{-1}\,e^{-i\pi/3}
- b_1   \,e^{i\pi/3}\,,
\eqa
then
\bqa
\left\{
\begin{tabular}{llll}
$T^+(q_1)$ =& $+b_{-2}            $ & $+ b_0       $ & $ + b_2$      \\
$T^+(q_2)$ =& $-b_{-2}            $ & $+ b_0       $ & $ - b_2$      \\
$T^0(q_3)$ =& $+b_{-2}e^{-2i\pi/3}$ & $+ b_0       $ & $ +b_2 e^{2i\pi/3}$
\end{tabular} \right.\,.
\eqa
This system is analogous to the system in \eqn{eq:evenc} with
the replacements $ c_i \to b_i$
and $T^+(q_3) \to T^0(q_3)$. Therefore its solution can be
directly read from
\eqn{eq:sysc}:
\bqa
 b_0       &=& \frac{T^+(q_1)+T^+(q_2)}{2}\,, \nl
 b_{\pm 2} &=& \left[
 \frac{T^+(q_1)-T^+(q_2)}{2}
 -e^{\pm 2 i \pi/3}(T^0(q_3)-b_0)
\right]
\frac{1}{1-e^{\mp 2 i \pi/3}}\,.
\eqa
Finally, from \eqn{eq:bq0} one obtains the last coefficient
\bqa \tld{b}_{00}(01)= \frac{R^{\prime \prime}(q_0)-b_0 - \sqrt{3}
\beta_0}{2 k_1^4 F}\,. \eqa

\end{document}